\documentclass[reprint, superscriptaddress]{revtex4-1}
\usepackage{geometry}
\usepackage{natbib}
\usepackage{graphics}
\usepackage{float}
\usepackage{setspace}
\usepackage{graphicx}
\usepackage{amsmath}
\usepackage{enumitem}
\usepackage{bm}
\usepackage{float}
\usepackage[latin1]{inputenc}
\usepackage{braket}
\usepackage{color}
\usepackage{verbatim}
\usepackage{bm}
\usepackage{bm,amssymb}
\usepackage{color}
\usepackage{placeins}
\usepackage{subfigure}
\usepackage{comment}

\newcommand{\Lyx}{L\kern-.1667em\lower.25em\hbox{y}\kern-.125emX\spacefactor1000}
\graphicspath{{820PNGSandPDFS/}}
\begin{document}
\bibliographystyle{plain} 
\pagestyle{plain} 
\pagenumbering{arabic}


\title{Divergence of the quadrupole-strain susceptibility of YbRu$_2$Ge$_2$; a local moment realization of electronic nematicity}
\author{E. Rosenberg}
\address{Department of Applied Physics, Stanford University}
\author{J.-H. Chu}
\affiliation{Department of Physics, University of Washington}
\author{J.C. Ruff}
\affiliation{Cornell High Energy Synchrotron Source, Cornell
University, Ithaca, NY 14853}
\author{A.T. Hristov}
\affiliation{Department of Physics, Stanford University}
\author{I.R. Fisher}
\address{Department of Applied Physics, Stanford University}

\maketitle
\textbf{
Ferroquadrupole order associated with local $4f$ atomic orbitals of rare earth ions is a realization of electronic nematic order. However, there are relatively few examples of intermetallic materials which exhibit continuous ferroquadrupole phase transitions, motivating the search for additional materials that fall in to this category. Furthermore, it is not clear \textit{a priori} whether experimental approaches based on transport measurements which have been successfully used to probe the nematic susceptibility in materials such as the Fe-based superconductors, will be as effective in the case of $4f$ intermetallic materials, for which the important electronic degrees of freedom are local rather than itinerant and are consequently less strongly coupled to the charge-carrying quasiparticles near the Fermi energy. In the present work, we demonstrate that the intermetallic compound YbRu$_2$Ge$_2$ exhibits a tetragonal-to-orthorhombic phase transition consistent with ferroquadrupole order of the Yb ions, and go on to show that elastoresistivity measurements can indeed provide a clear window on the diverging nematic susceptibility in this system. This material provides a new arena in which to study the causes and consequences of electronic nematicity. 
}

\section{Introduction}
Electronic nematic order in crystalline solids corresponds to a spontaneously broken discrete rotational symmetry driven by interactions between low-energy electronic degrees of freedom \cite{SKreview}. The empirical observation that nematic order occurs in, or near to, superconducting phases in Fe-based materials \cite{add2, add3, ChuScience2012, fernandesNatPhys2014, add6} and possibly also cuprates \cite{cuprate1, cuprate2, cuprate3, cuprate4, cuprate5, cuprate6, cuprate7, cuprate8, cuprate9} motivates a series of fundamental questions about the possible role(s) that nematicity might play in such systems \cite{addnematicity1, addnematicity2, Vojta2009, lawler2010intra}. Fe-based and cuprate superconductors are complicated materials, with a variety of other intertwined electronic phases, making it highly desirable to identify simpler model systems that exhibit nematic order for which the underlying effective Hamiltonian is better-understood. Ferroquadrupole order associated with local $4f$ atomic orbitals of rare earth ions is a realization of electronic nematic order \cite{TransIs}, and in principle can provide just such a model system.

The effective Hamiltionian describing the low energy properties of rare earth ions incorporated in crystalline solids is understood in great detail. The $4f$ electronic wavefunctions are spatially localized, and strong spin-orbit coupling yields an electronic multiplet that is described by a total angular momentum number $J$. Due to the local nature of the electrons, the crystal electric field (CEF) from the surrounding ligands acts as a weak perturbation,  splitting the degeneracy of the $2J+1$ spherical harmonic basis states. Significantly, the low energy states of these $4f$ electronic multiplets can exhibit degeneracies or near degeneracies. In these degenerate or nearly degenerate manifolds, there can exist states with large multipolar moments. Both the coupling of the lattice with quadrupolar degrees of freedom, and a $4f-4f$ coupling from the generalization of the RKKY interaction mediated by conduction electrons, can then provide energetically favorable conditions for multipolar instabilities to occur \cite{Morinreview}. A variety of multipolar behavior has been observed in such materials. The specific case of ferroquadrupole order, in which each atomic site takes on a quadrupole moment of the same orientation, breaks the rotational symmetry of the point group of the crystal and hence provides an example of a lattice in which sites exhibit collective nematic order. There are, however, relatively few examples of intermetallic materials which exhibit continuous ferroquadrupole phase transitions. The primary examples are tetragonal-to-orthorhombic phase transitions observed for TmAg$_2$ \cite{MorinTmAg2} and TmAu$_2$ \cite{MorinTmAu2, KosakaTmAu2}, motivating the search for additional materials that fall in to this category. The present work adds YbRu$_2$Ge$_2$ to that list.

YbRu$_2$Ge$_2$ has a tetragonal crystal structure at room temperature (space group I4/mmm), belonging to the same ThCr$_2$Si$_2$ structure type as the familiar ``122" Fe-pnictides such as BaFe$_2$As$_2$. The $4f$ electrons of the trivalent Yb ion form local multiplets with a total angular momentum quantum number J = 7/2. The CEF for the tetragonal point group symmetry results in 4 Kramers doublets, two with $\Gamma_6$ and two with $\Gamma_7$ character \cite{YRGTheory} (see Appendix~\ref{section:S1}). The material displays some weak heavy fermion characteristics \cite{YRGCEF}, but the low temperature properties can be well-understood in terms of these local $4f$ orbitals \cite{YRGTheory}. Analysis of heat capacity, susceptibility and inelastic neutron scattering measurements indicate that the CEF groundstate is a pseudo-quartet comprising states with almost pure $|\pm1/2\rangle$ ($\Gamma_6$) and $|\pm3/2\rangle$ ($\Gamma_7$) character, split by approximately 300 K from the other two doublets \cite{YRGCEF,YRGTheory}.   The material exhibits three continuous phase transitions at low temperature associated with the pseudo-quartet: a non-magnetic phase transition at $T_Q$ = 10.2K, and two magnetic phase transitions (one at $T_{N1}$ = 6.5K, corresponding to the onset of a collinear amplitude modulated antiferromagnet state, and a subsequent one at $T_{N2}$ = 5.7K, the origin of which is currently unknown  \cite{YRGCEF,MuonYRG}). Based on analysis of the multipole moments of the CEF groundstate, combined with a mean-field treatment of the effective Hamiltonian for the system, it has been suggested that the origin of the non-magnetic transition is a ferroquadrupolar ordering of the local $4f$ electrons (see Ref. \cite{YRGCEF} and Appendix~\ref{section:S3}). However, to-date no measurements have confirmed the ferroquadrupolar nature of the phase transition, nor has the quadrupole strain susceptibility been investigated. 

In the present work,  we use high-resolution X-ray diffraction to reveal that the non-magnetic phase transition corresponds to a tetragonal-to-orthorhombic phase transition, consistent with spontaneous ferroquadrupole order with a B$_{1g}$ symmetry. We then present elastoresistivity measurements which reveal a divergence of the quadrupole strain susceptibility (i.e. the nematic susceptibility) in the same (B$_{1g}$) symmetry channel, $\chi_{B_{1g}}$. Comparison of the divergence of $\chi_{B_{1g}}$ and the critical temperature for the ferroquadrupole order $T_Q$ allows evaluation of the Levy criterion (defined in Section II), from which it is inferred that the ferroquadrupole transition in YbRu$_2$Ge$_2$ is driven primarily by magneto-elastic coupling. Our measurements establish YbRu$_2$Ge$_2$ as a model system to explore electronic nematicity, for example as a possible starting point from which the ferroquadrupole order could be continuously suppressed towards a nematic quantum phase transition.

\begin{figure*}
\centering
\includegraphics[width=0.95\textwidth]{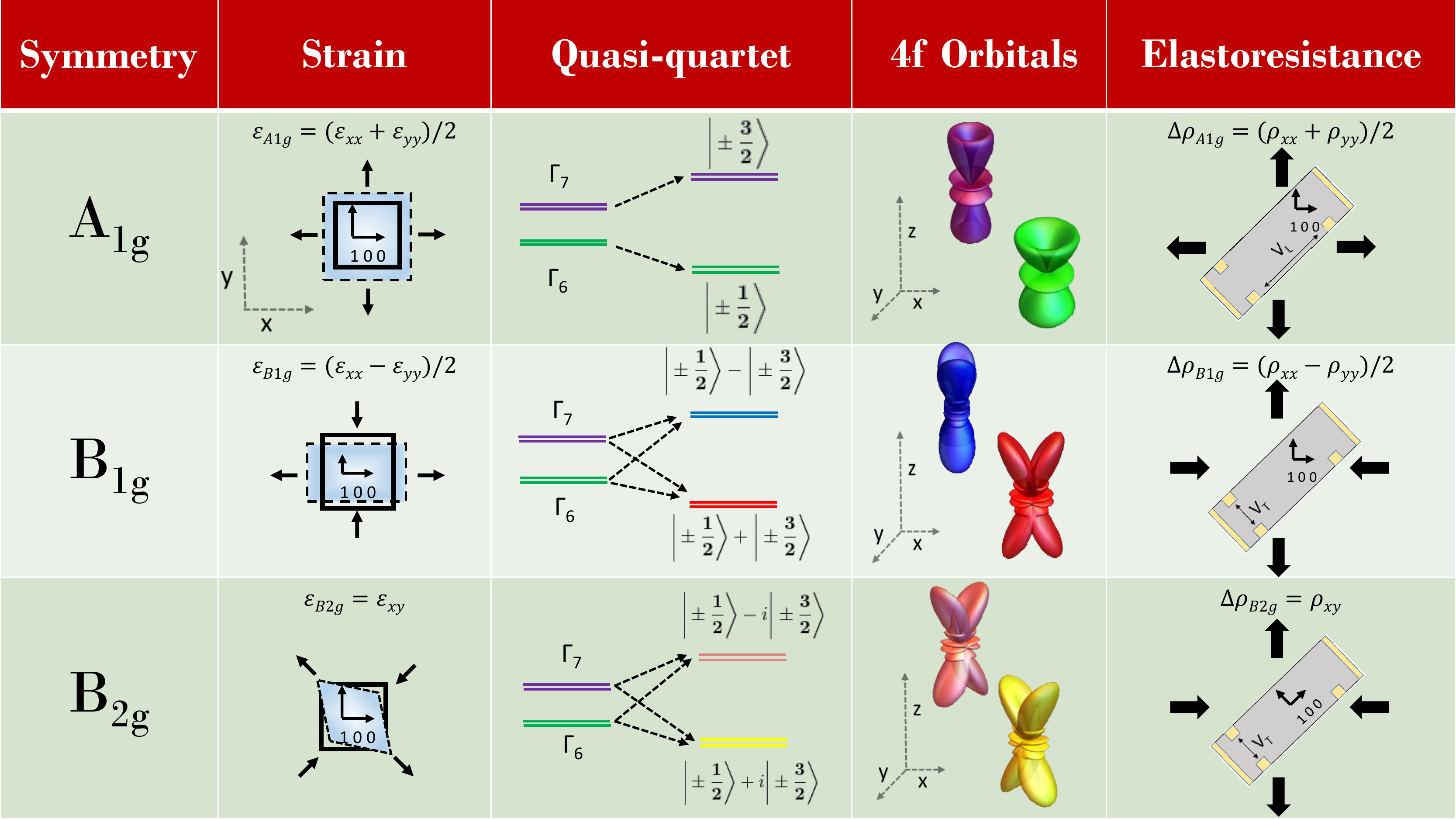}

\caption{\label{fig1} \textbf{Effect of strains of different symmetries on the CEF eigenstates of YbRu$_2$Ge$_2$, and measurement of the associated elastoresistivity coefficients.} Strains corresponding to distinct irreducible representations, induced by external stresses, are indicated by thick black arrows. These strains admix and split the CEF eigenstates; the effect on the low energy CEF quasi-quartet and the charge density of the ground state orbitals is illustrated. For simplicity, the $\Gamma_6$ and $\Gamma_7$ eigenstates are illustrated for pure $|\pm1/2\rangle$ and $|\pm3/2\rangle$ character respectively (see main text). The far right column shows the required orientation of the normal strain (thick black arrows), crystal axes (insets to grey schematic crystals) and voltage contacts (yellow pads, with measured voltages indicated for longitudinal and transverse pairs) for measurement of the associated elastoresistivity coefficients in each symmetry channel.
}
\end{figure*}

\section{Quadrupole Strain Susceptibility and the Levy Criterion}
We start by describing the quadrupole strain susceptibility (i.e. the nematic susceptibility for the $4f$ quadrupolar system) and its significance for ferroquadrupole order. 

A minimal Hamiltonian can be written down which describes the quadrupolar behavior of the $4f$ electrons in YbRu$_2$Ge$_2$:

\begin{equation}
 H_{eff}=H_{CEF}+\sum_i{-\bigg(B_i Q_i \varepsilon_i + K_i \langle Q_i \rangle Q_i}\bigg)
\end{equation}

 The first term in the sum is the bilinear magnetoelastic quadrupole-strain coupling, and the second term is a mean-field quadrupole interaction term originating from an effective RKKY coupling mediated by the conduction electrons \cite{Morinreview}. 
 The sum is over the three relevant symmetry channels for the tetragonal system, A$_{1g}$, B$_{1g}$, and B$_{2g}$. $Q_i$ are the quadrupole operators of the $i$th symmetry channel (for B$_{1g}$ symmetry these are O$_2^2$ = J$_x^2$-J$_y^2$, for B$_{2g}$ symmetry P$_{xy}$ = (J$_x$J$_y$+J$_y$J$_x$)/2), and $\varepsilon_i$ is the corresponding strain (for B$_{1g}$ symmetry, ($\varepsilon_{xx}-\varepsilon_{yy}$)/2, for B$_{2g}$ symmetry, $\varepsilon_{xy}$). H$_{CEF}$ describes the crystal electric field from the surrounding Ru and Ge atoms that produces the ground state quasi-quartet (See Appendix~\ref{section:S1}). The gap between the ground state doublet and first excited doublet has been proposed to be $\Delta_{CEF_1}\approx 10K$ from entropy considerations \cite{YRGCEF}. The other doublets have been shown to be much higher in energy ($\Delta_{CEF_2}\approx 300K$) and can be ignored for any low temperature physics.

Strain $\epsilon_i$, induced by an external stress, admixes and splits the CEF eigenstates (see Figure 1), resulting in an induced quadrupole moment. The associated quadrupole strain susceptibility $\chi_i$ can be defined for each symmetry channel $i$. Considering first the case in which $K_i$ = 0, the quadrupole strain susceptibility has the form
\begin{equation}
\chi_{i}=\frac{dQ_{i}}{d\varepsilon_{i}}\bigg|_{\varepsilon\to 0}= \frac{2\langle q_{i}\rangle_{o}^2 B_{i}}{\Delta_{CEF_{1}}}\tanh{\bigg(\frac{\Delta_{CEF_{1}}}{2T}\bigg)}
\end{equation}
 where $\langle q_i \rangle_o$ is the magnitude of the quadrupole moment (see Appendix~\ref{section:S3}). In the limit $T > \Delta_{CEF_1}$, this reduces to an approximate Curie form $\chi_i \sim \frac{\langle q_i \rangle_{o}^2B_i}{T}$, which becomes exact for $\Delta_{CEF_1} = 0$.

Adding quadrupole interactions to this model renormalizes the quadrupole-strain susceptibilities. Considering again the limit $\Delta_{CEF_1}=0$, this results in a Curie-Weiss functional form: 
\begin{equation}
\chi'_{i}=\bigg ( \frac{dQ}{d\epsilon}\bigg )_i = \frac{\langle q_i \rangle_{o}^2 B_i}{T-\langle q_i \rangle_{o}K_i}
\end{equation}

Thus the Weiss temperature of the quadrupole-strain susceptibility in a given symmetry channel is proportional to the strength of the quadrupole-quadrupole interaction term and the saturated quadrupole moment in that channel. If the lattice were infinitely stiff (i.e. strain is treated as a fixed parameter), the system would have a tendency to undergo quadrupolar order at this ``bare" critical temperature.

Of course, the lattice is not infinitely stiff, and the magneto-elastic coupling renormalizes the critical temperature. Minimizing the free energy with respect to strain,  including the elastic energy terms, the ordering temperature is found to be:

 \begin{equation}
T_Q=\bigg(\big(K+\frac{B^2}{C}\big)\langle q\rangle_{o}\bigg)_i=G_i\langle q_i\rangle_{o}
\end{equation}

Where B and C are the magnetoelastic and elastic constants respectively of the ordering symmetry channel i, and G represents the renormalized quadrupolar interaction coefficient.

The relative magnitudes of $\frac{B_i^2}{C_i}$ and $K_i$ dictates whether the nematic phase transition is driven by strain, in the traditional Cooperative Jahn-Teller sense \cite{GehringGehring} or by electronic quadrupolar fluctuations, as inferred for certain $4f$ ferroquadrupole systems \cite{Morinreview, MorinTmAg2, KosakaTmAu2} and underdoped ``122" Fe-pnictides \cite{ER, ChuScience2012, add6}. Thus a precise measurement that determines the ratio of these two components, known as the Levy criterion, is an essential development to understanding the nature of quadrupolar phase transitions in any given $4f$ system.

While the quadrupole-strain susceptibility is frequently invoked from a theoretical perspective \cite{Morinreview}, it has rarely been determined experimentally. Simple consideration of thermodynamic relations imply that $\chi_i$ can be inferred from softening of the elastic constants, magnetostriction, and measurements of the third-order magnetic susceptibility \cite{Morinreview}, and all of these approaches have been utililized previously in other candidate $4f$ systems \cite{MorinTmAu2}. However, while measurements of the elastic moduli provide a direct perspective on the lattice symmetry modes, a large temperature dependent background and a strict reliance on a suitable geometry of the sample make it difficult in practice to obtain the temperature-dependence of $\chi_i$ with quantitative accuracy. Similarly, magnetic measurements face difficulties in obtaining the temperature-dependence of $\chi_i$, in part because of issues with demagnetization fields, and in part because quadrupolar-field susceptibilities  must be untangled from the quadrupole-strain susceptibilities to properly infer them. Here, rather than inferring $\chi_i$ based on a combination of thermodynamic measurements, we outline how this quantity can be determined via elastoresistance measurements.

\section{Correspondence between elastoresistivity and quadrupole-strain susceptibility}

Although resistivity is not a thermodynamic measurement, it is a sensitive measure of the electronic scattering rate in a crystal. Considering $4f$ intermetallic systems, the local $4f$ electronic sites serve as strong scatterers of the conduction electrons. Anisotropic scattering from these sites and their associated quadrupolar moments has previously been described for cubic materials \cite{Anisores} but not investigated in the context of quadrupolar phase transitions in tetragonal materials. Under two conditions, which can be readily justified for YbRu$_2$Ge$_2$, we can directly relate a change in the normalized anisotropic resistivity (divided by the isotropic resistivity) to a change in the corresponding quadrupole moment (see Appendix~\ref{section:S5} and \cite{Aniso}). The first condition is that the resistivity is dominated by scattering of the conduction electrons from the lattice of $4f$ electronic mutliplets. As we demonstrate, this is indeed the case for YbRu$_2$Ge$_2$. The second conditon is that the induced quadrupole moment is perturbative, which is by definition the case for the technique that we use to determine the elastoresistivity. 

When these two conditions are satisfied, an elastoresistivity technique, in which a tunable strain is applied to a material while measuring the induced anisotropic resistivity (illustrated schematically in Figure 1), provides an ideal method to directly probe the temperature-dependence of the material's quadrupole-strain susceptibility. For example, considering the B$_{1g}$ symmetry channel (second row of Figure 1), a B$_{1g}$ symmetry strain (caused by some external stresses) induces a finite B$_{1g}$ symmetry change in the resistivity $(\frac{\Delta\rho}{\rho})_{B_{1g}} = \frac{(\rho_{xx}-\rho_{yy})}{\rho_0}$ where $\rho_{xx}$ and $\rho_{yy}$ are measured in the presence of the strain and $\rho_0$ is the unstrained isotropic in-plane resistivity. If the scattering rate $\Gamma$ is dominated by scattering from the 4f orbitals, and if the effective mass of the conduction electrons is not strongly affected by the induced strain, then
\begin{equation}
\bigg(\frac{\Delta \rho}{\rho}\bigg)_{B_{1g}} \approx \bigg(\frac{\Gamma_{xx}-\Gamma_{yy}}{{(\Gamma_{xx}})_0+({\Gamma_{yy}})_0}\bigg)_{4f} \propto \langle O^2_2 \rangle
\end{equation}

Taking the appropriate derivatives with respect to strain gives:
\begin{equation}
\cfrac{\bigg(\cfrac{d\rho}{\rho}\bigg)_{B_{1g}}}{d\varepsilon_ {_{B_{1g}}}}   \propto {\chi}_ {_{B_{1g}}}
\end{equation}

Thus, measuring the symmetry decomposed elements of the elastoresistivity tensor provides direct measurements of the corresponding quadrupole-strain susceptibilities. Appropriate orientation of crystal axes, normal strain axes, and contact placement necessary to measure each symmetry channel are illustrated in Figure 1, and described in greater detail in \cite{Max}.

Elastoresistance techniques have previously been used to determine the nematic susceptibility of materials for which the nematicity derives from itinerant electronic states, including Fe-based superconductors \cite{ER, addER1, addER2, Max, Nonlin, Alex} and URu$_2$Si$_2$ \cite{URS}. Here, we demonstrate that the technique is also appropriate to determine $\chi_i$ of materials that undergo ferroquadrupole order of local $4f$ orbitals.  Since for such systems $\chi_i$ can in principle be large in several symmetry channels $i$, we introduce new means (described in Section IV) to ensure appropriate symmetry decomposition, using a Focused Ion Beam to prepare the sample for each measurement.

\section{Experimental Methods}

Single crystals of YbRu$_2$Ge$_2$ were grown from a high-temperature indium flux, as described elsewhere \cite{Thesis} (see Appendix~\ref{section:S9}). The flux was decanted in a centrifuge and the crystals etched in HCl acid for several months. The resulting crystals, which had a plate-like morphology, with the c-axis perpendicular to the plane of the plates, could be cleaved to a thickness of 20$\mu$m or less, and were identified and oriented using single crystal X-ray diffraction.

Low temperature, high-resolution, X-ray diffraction measurements were performed on beamline A2 at CHESS (Cornell High Energy Synchrotron Source). The intensity profiles of the (6 0 0) Bragg peak were mapped out in momentum space from 40K to 6.6K.  The orthorhombicity parameter $(a-b)/(a+b)$ was obtained by finding the first moment of the intensity (integrating counts multiplied by position) of line scans of the orthorhombic Bragg peaks (see Appendix~\ref{section:S2})

Flat samples with a uniform thickness between 5 and 20 $\mu$m  were bonded to 50 $\mu$m thick Si substrates using Angstrom Bond (AB9112-2.5G). Current and voltage contacts were made by connecting gold wires to the sample using Epotech H-20E silver paste. Hall-bar patterns were then etched into the sample using a Focused Ion Beam (FIB) (see Appendix~\ref{section:S8} for beam parameters) in order to precisely define current directions and contact geometry (Figure 1). Typical bar dimensions were 200-400 $\mu$m in length, by 70-120 $\mu$m in width. The Si substrate was then bonded to the center of a side of a Piezomechanick 5x5x9 mm PZT piezoelectric stack (Piezomechanik PSt150/5x5/7 cryo 1) in the appropriate orientation using Devcon 5-Minute Epoxy. Uncertainty in alignment with respect to the crystal axes was estimated to be less than 2 degrees.  Strains experienced by the sample as a consequence of varying the voltage applied to the PZT stack were estimated via resistive strain gauges positioned on the opposite side of the piezoelectric stack, as well as a resistive strain gauge glued on top of a Si substrate mounted on the PZT \cite{ER}. The PZT stack was mounted on a thermally anchored probe in a helium flow cryostat. 

To ensure that the elastoresistive response was linear, the voltage applied to the PZT stack was varied while the temperature was held constant, and the sample's response plotted against that of the strain gauge. For both B$_{1g}$ and B$_{2g}$ responses, the elastoresistance was always linear. Detailed elastoresistivity measurements were then taken using an amplitude demodulation technique, following the method described in Ref. \cite{Alex}. This technique was applied using a 1.6Hz 50V rms sinusoidal excitation while simultaneously driving a 5 mA rms current through the sample at 107 Hz, sourced by a Keithley 6221 DC and AC current source. The combination of mechanical and electrical excitation produces voltage signals at the sum and difference of the two frequencies, proportional to the unstrained resistivity of the sample. The signal is first measured by a SRS830 lock-in amplifier (LIA) with a time constant $<$ 30ms referenced to 107 Hz, and the output of that LIA is measured by a second LIA with a time constant of $>3s$ referenced at 1.6Hz to detect the elastoresisitivity signal. Care was taken to ensure the phase of the output signal was properly accounted for and the strain was measured simultaneously \cite{Alex}. Strain transmission was measured from the piezo surface to the top surface of the Si where the samples were glued, and was determined to be roughly 50\% and essentially temperature independent from 4K to 100K.  This was taken into account in determining the magnitude of the elastoresistivity coefficients.   Data were taken while the temperature was slowly swept between 4 K and 100 K. 


\begin{figure}
\centering
\includegraphics[width=1\linewidth]{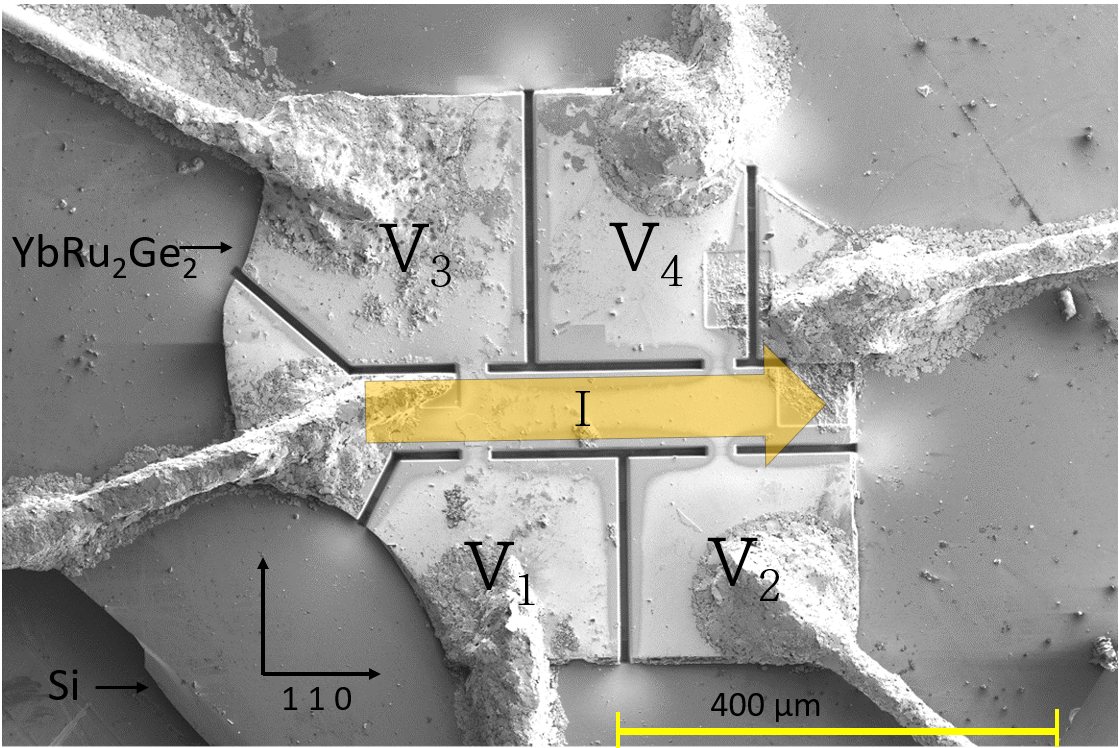}

\caption{\label{fig2}\textbf{Scanning electron micrograph of a representative micro-structured sample used for elastoresistance measurements} The YbRu$_2$Ge$_2$ sample is glued to a 50 $\mu$m thick Si substrate.The crystal axes (indicated by black arrows) were previously determined by x-ray diffraction, and identified here by the horizontal facets. Line cuts were etched using a Focused Ion Beam to control the current path (indicated by yellow arrow) and precisely define the voltage contact positions (labelled V$_{1}$ through V$_{4}$) .}
\end{figure}
\section{Results and Discussion}

\subsection{High Resolution X-ray Diffraction}
Splitting of the (6 0 0) Bragg peak was observed below 10.2K along the (1 1 0) and (1 -1 0) directions, with the new peaks indicative of a $B_{1g}$ orthorhombic structural distortion. The temperature dependence of the orthorhombicity parameter of YbRu$_2$Ge$_2$ is shown in Figure 3, together with the anticipated temperature dependence of a mean-field Ising order parameter that onsets at 10.2 K.  No evidence for a similar phase transition is found for the non-magnetic analog YRu$_2$Ge$_2$ (see \cite{YRuGe} and inset to Figure 4). The data for YbRu$_2$Ge$_2$ are consistent with a continuous ferroquadrupole phase transition, as predicted previously based on analysis of the CEF parameters \cite{YRGCEF}. Of the two possibilities (corresponding to either $B_{1g}$ or $B_{2g}$ symmetry), these measurements detemine that the ferroquadrupolar state has a $B_{1g}$ symmetry.
\begin{figure}[H]
\includegraphics[width=0.98\linewidth]{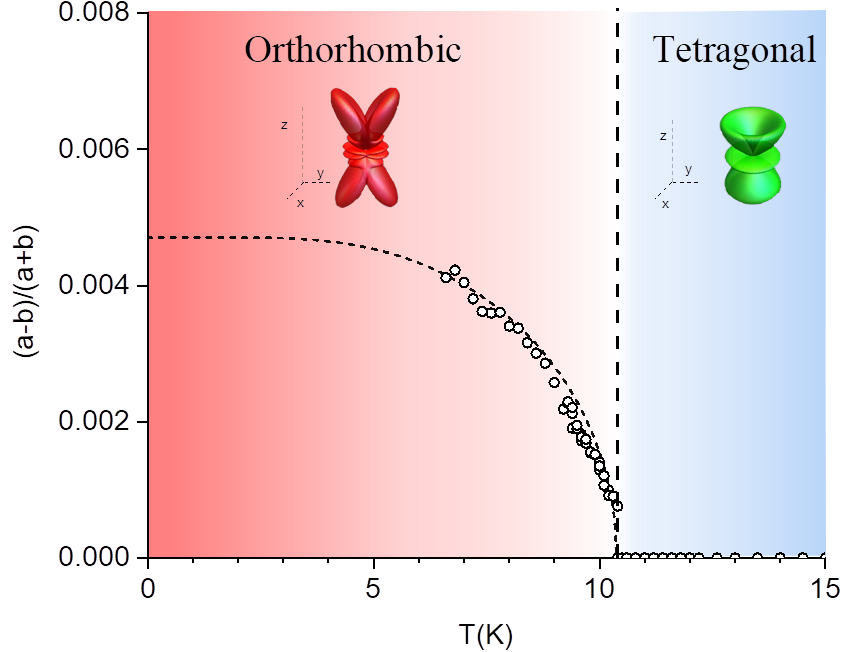}

\caption{\label{fig1} \textbf{Orthorhombicity of YbRu$_2$Ge$_2$} The order parameter of the orthorhombic structural phase transition is plotted against temperature. A mean-field Ising order parameter that becomes finite is 10.2K is overlaid as a guide to the eye. Insets illustrate the charge density of the $4f$ CEF quasi-quartet groundstate of YbRu$_2$Ge$_2$ in the tetragonal phase (green, with 4-fold symmetry), and that of the orthorhombic phase (red, with two-fold symmetry) . Ferroquadrupole order of the $4f$ orbitals is inferred from the observed orthorhombicity.}
\end{figure}

\subsection{Resistivity}
\begin{figure}[H]
\centering
\includegraphics[width=1\linewidth]{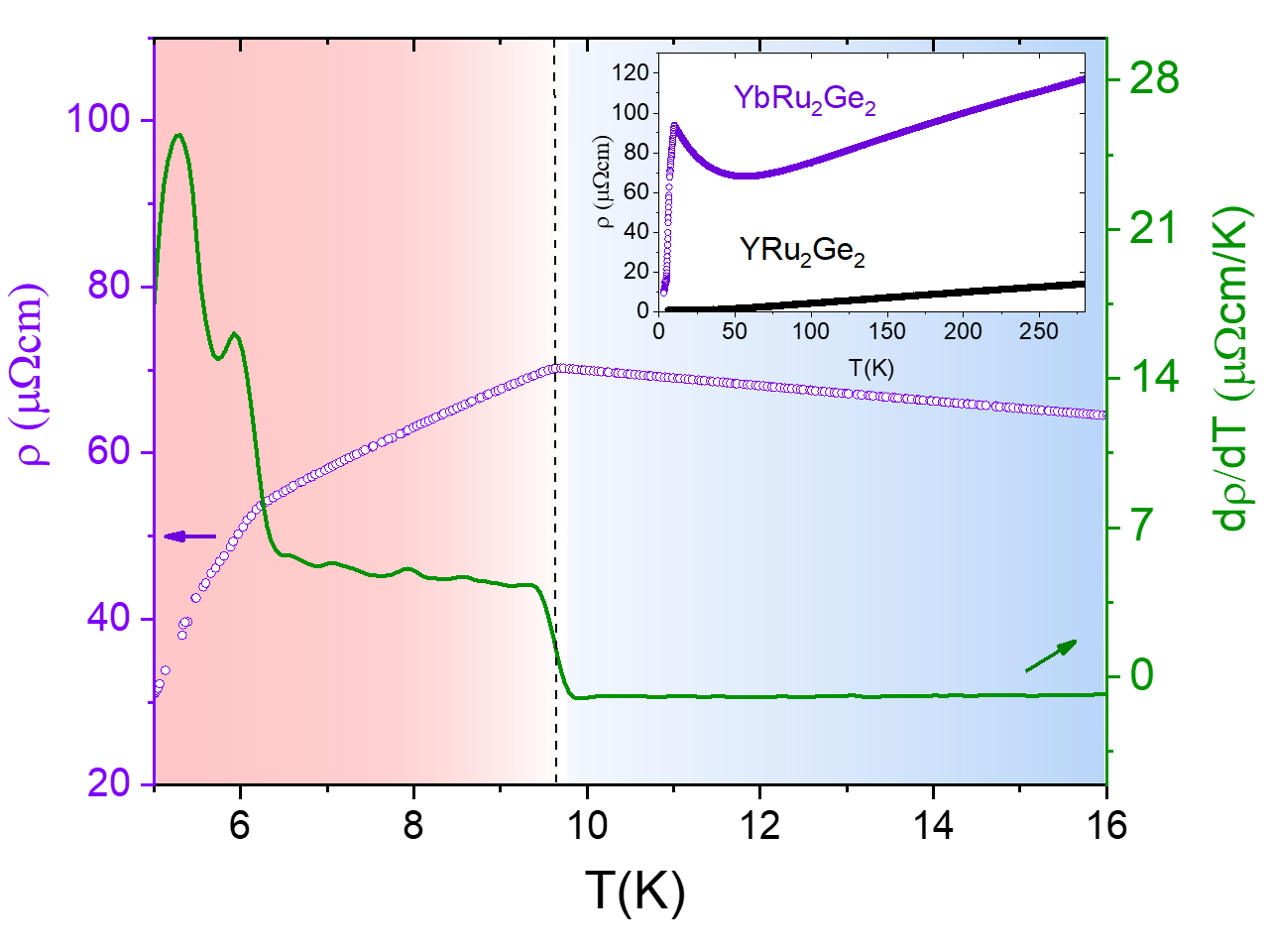}

\caption{\label{fig1} \textbf{Temperature dependence of the resistivity of YbRu$_2$Ge$_2$.} The resistivity (purple data points) and its derivative with respect to temperature (green data points) show clear features at $T_Q$ (marked by vertical dashed line), as well as $T_{N1}$ and $T_{N2}$. Slight differences in estimates of the critical temperatures relative to values extracted from x-ray diffraction measurements are attributed to differences in sweep rates of the two measurements. Inset shows the resistivity over a larger temperature range, in which a different YbRu$_2$Ge$_2$ single crystal is measured.  Data are also shown for the non-magnetic analog, YRu$_2$Ge$_2$, for which no phase transitions are observed. Yttrium has an empty $4f$ shell but otherwise the electronic structure of the two materials is anticipated to be very similar, and differences in the resistivity between the two compounds are primarily attributed to $4f$ scattering.}

\label{fig:Res}
\end{figure}

The temperature dependence of the in-plane resistivity of unstrained (i.e. free-standing) YbRu$_2$Ge$_2$ is shown in Figure~\ref{fig:Res}. The data are consistent with previously published measurements, revealing a Kondo-like rise below 50 K and sharp features (seen more clearly in the derivative $\frac{d\rho}{dT}$) signifying the ferroquadrupole and magnetic phase transitions. Similar measurements of the non-magnetic iso-structural, iso-electronic analog, YRu$_2$Ge$_2$, also shown in Figure~\ref{fig:Res}, reveal a considerably smaller in-plane resistivity, and no signatures of any phase transitions. Since the band structure of the two materials is presumably quite similar, the large difference in the resistivity can be attributed to scattering from the local $4f$ orbitals in YbRu$_2$Ge$_2$. Significantly, since the resistivity of YbRu$_2$Ge$_2$ in the temperature window above $T_Q$ (blue shading in Figure 4) is dominated by $4f$ scattering, the elastoresistivity (described below) can be directly related to the quadrupole strain susceptibility in the same symmetry channels, as described in Section III.

\subsection{Linear B$_{1g}$ and B$_{2g}$ Elastoresistivity}

Elastoresistivity measurements were performed for multiple samples. Responses in the B$_{1g}$ and B$_{2g}$ symmetry channels were always found to be linear. Representative data for both of these symmetry channels (taken for samples of equal dimensions) are shown in Fig 5 as a function of temperature. Elastoresistivity data for the A$_{1g}$ symmetry channel were also recorded during these measurements; while these data do not correspond to a susceptibility of a symmetry breaking quadrupole order, the measurements are presented in Appendix~\ref{section:S7} for completeness. The response in the B$_{1g}$ channel was found to be considerably larger than that in the B$_{2g}$ channel. If the proportionality constants relating the resistivity anisotropy to the local quadrupole moment (see Appendix~\ref{section:S5}) are similar for the two symmetry channels, which would be the case for an isotropic Fermi surface but is certainly not guaranteed by symmetry, then the observed large anisotropy in the elastoresistivity would imply that the magneto-elastic coupling coefficient B$_i$ for $i=B_{1g}$ is significantly larger than that for the B$_{2g}$ symmetry channel. This would certainly be consistent with the observed B$_{1g}$ symmetry of the orthorhombic distortion, but needs to be verified by other thermodynamic measurements. The B$_{1g}$ and B$_{2g}$ elastoresistivity coefficients were also measured for YRu$_2$Ge$_2$. No significant temperature dependence for either channel was found within experimental error, as anticipated given the absence of partially filled $4f$ orbitals in this compound.

Since the resistivity of YbRu$_2$Ge$_2$ in this temperature range is dominated by scattering from $4f$ orbitals (Figure 4), the temperature dependence of the elastoresistivity provides a good measure of the temperature dependence of the quadrupole strain susceptibility. Motivated by the discussion in Section II, the B$_{1g}$ elastoresistivity was fit to a Curie-Weiss functional form:

\begin{equation}
\chi_{B_{1g}}^{fit}=-\frac{448.6\pm1.2}{T-0.11\pm0.03} +4.48 \pm 0.02
\end{equation}

The quality of fit over the full temperature range, right down to T$_Q$, (Figure 5 and inset to Figure 5), indicates that the splitting $\Delta_{CEF_1}$ of the quasi-quartet in the tetragonal state is less than $T_Q$ in magnitude, consistent with previous estimates from inelastic neutron scattering \cite{MuonYRG}. Significantly, with reference to Equation (2), since the Weiss temperature is found to be close to 0K, it is clear that quadrupolar interactions mediated by the conduction electrons do not play an important role in the eventual ferroquadrupole phase transition. In other words, referring back to Equation (4) and the Levy criterion, we find that $B_i^2/C_i >> K_i$ for YbRu$_2$Ge$_2$ in the B$_{1g}$ symmetry channel, and hence deduce that the ferroquadrupole transition must be driven primarily by magneto-elastic coupling in this material.

\begin{figure}
\centering
\includegraphics[width=1\linewidth]{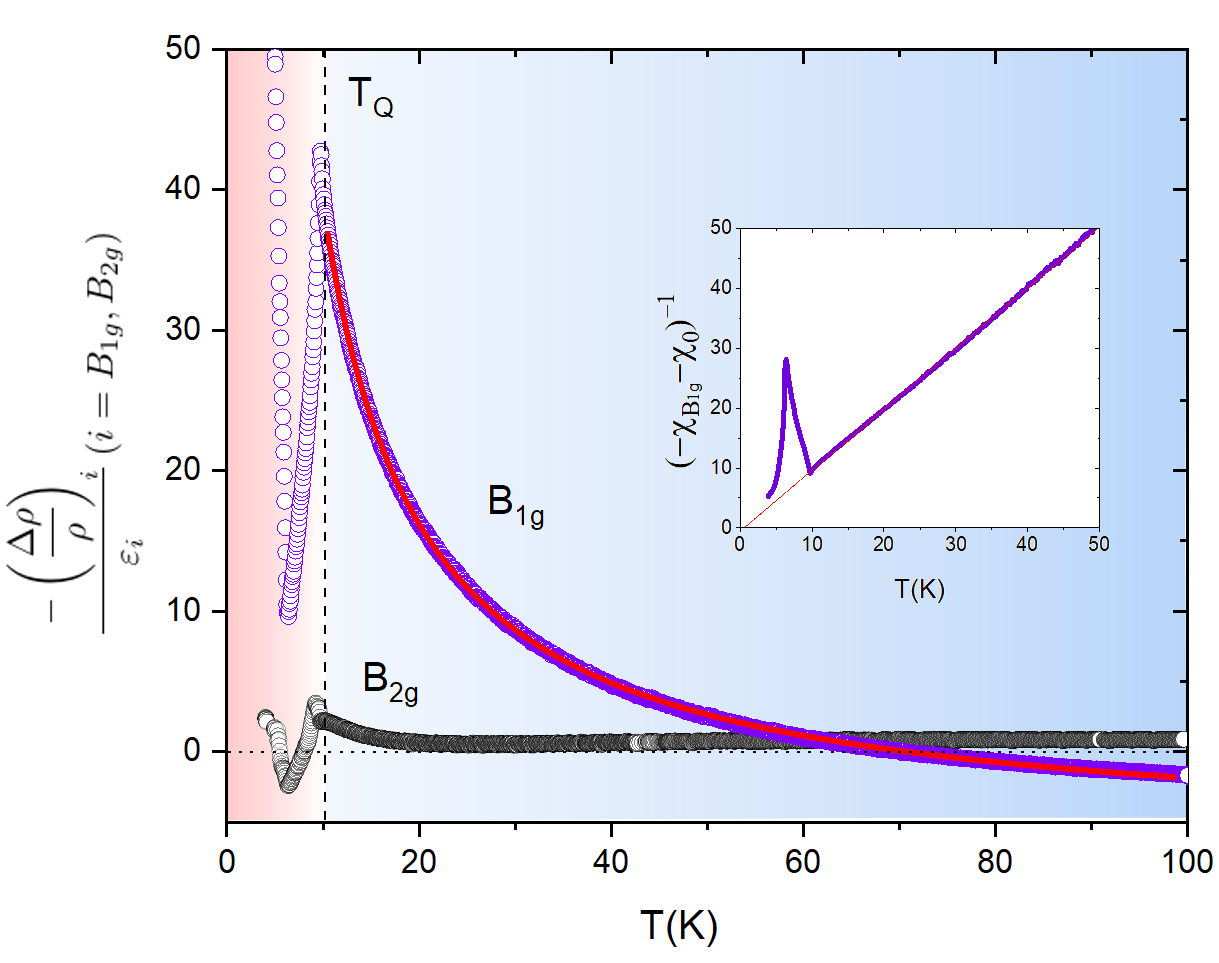}

\caption{\label{fig3}\textbf{Temperature-dependence of the elastoresistivity of YbRu$_2$Ge$_2$ in the B$_{1g}$ and B$_{2g}$ symmetry channels.} As described in the main text, these quantities (shown by blue and black data points respectively) are proportional to the quadrupole strain susceptibility of the same symmetry channels. Inset shows the inverse of the quadrupole strain susceptibility $ (-\chi_{B_{1g}} - \chi_0)^{-1}$ for the B$_{1g}$ channel having subtracted a temperature-independent offset $\chi_0 = 4.48$. Red lines in the main panel and inset show fit to a Curie-Weiss functional form, with fit parameters described in the main text. The Weiss temperature is found to be 0 K within experimental error. Vertical dashed lines indicate $T_Q$, separating tetragonal (blue) from orthorhombic (red) phases. }
\end{figure}

\section{Implications for the Electronic Nematicity in Y\MakeLowercase{b}R\MakeLowercase{u}$_2$G\MakeLowercase{e}$_2$}

YbRu$_2$Ge$_2$ is a counterpoint to underdoped Fe-based superconductors, such as BaFe$_2$As$_2$, in terms of how the nematic phase transition is driven. Numerous measurements, including elastoresistivity \cite{ER, addER1, addER2, Max, Nonlin, Alex}, elastic moduli \cite{FernElast, addelastic}, Raman scattering \cite{GallaisRaman, addRaman1, addRaman2, addRaman3} and NMR \cite{KissikovNMR, addNMR1}, have shown that the nematic susceptibility of BaFe$_2$As$_2$ and its analogs follow a Curie-Weiss functional form, with a Weiss temperature that is only slightly lower than the critical temperature for the coupled nematic/structural phase transition. Thus, while nematic-elastic coupling is still a necessary factor in driving the structural phase transition in those materials, the critical temperature is predominantly determined by electronic coupling between the nematic degrees of freedom \cite{Morinreview}. In contrast, YbRu$_2$Ge$_2$ presents a case in which magneto-elastic coupling is the dominant driving force for the phase transition. Even so, the material is clearly nematic: a separate degree of freedom (the local $4f$ quadrupole moment) that couples bi-linearly to elastic strain with the same symmetry is evident from the temperature dependence of the quadrupole strain susceptibility, and its divergence drives a pseudo-proper ferroelastic phase transition. The lattice provides the effective coupling between the local quadrupole moments, but does not itself harbor any tendency towards a structural instability, as demonstrated by the absence of a comparable phase transition in the non-magnetic analog YRu$_2$Ge$_2$.
 
The significance of the present observation is twofold. First, we have shown that elastoresistance measurements can be used to probe the nematic susceptibility (quadrupole strain susceptibility) for local $4f$ systems, at least in this specific case of YbRu$_2$Ge$_2$. 
This provides a new avenue to characterize a critically important characteristic of intermetallic quadrupolar systems. Since the quadrupole strain susceptibility can, at least in principle, be large in several symmetry channels, we adopted a new methodology of micro-structuring the crystal using a FIB to guarantee symmetry decomposition of the resulting elastoresistivity. Second, we have established YbRu$_2$Ge$_2$ as a model system to explore effects of nematicity in metals. These ideas and measurements can be extended in this system, for example probing nematicity proximate to a putative quantum phase transition if the ferroquadrupole order were to be suppressed towards zero temperature. Methods of suppressing ferroquadrupole order using doping, magnetic field, or transverse strain \cite{TransIs} are readily experimentally accessible.\\

\section{Acknowledgements}
We acknowledge fruitful conversations with J. Straquadine and S. Kivelson. EWR and IRF were supported by the
Gordon and Betty Moore Foundation Emergent Phenomena in Quantum
Systems Initiative through Grant GBMF4414. Part of this work was performed at the Stanford Nano Shared Facilities (SNSF), supported by the National Science Foundation under award ECCS-1542152.  J.-H. C. acknowledges the support from the State of Washington funded Clean Energy Institute. Research conducted at the Cornell High Energy Synchrotron Source (CHESS) is supported by the NSF \& NIH/NIGMS via NSF award DMR-1332208. A.T.H is supported by a NSF Graduate Research Fellowship under grant DGE-114747.

\bibliographystyle{plain} 


\onecolumngrid
\newpage
\appendix

\section*{Supporting Information (SI)}
\section{CEF splitting of the J=7/2 Hund's rule multiplet of YbRu$_2$Ge$_2$}
\label{section:S1}
The Hund's rule groundstate multiplet of the Yb$^{3+}$ ion, which is characterized by a total angular momentum J=7/2, is split by the crystal electric field (CEF) according to the effective Hamiltonian:

\begin{equation}
H_{CEF}=B_2^0O_2^0+B_4^0O_4^0+B_4^2O_4^4+B_6^0O_6^0+B_6^4O_6^4
\end{equation}

\begin{figure}[H]
\centering
\includegraphics[width=0.5\linewidth]{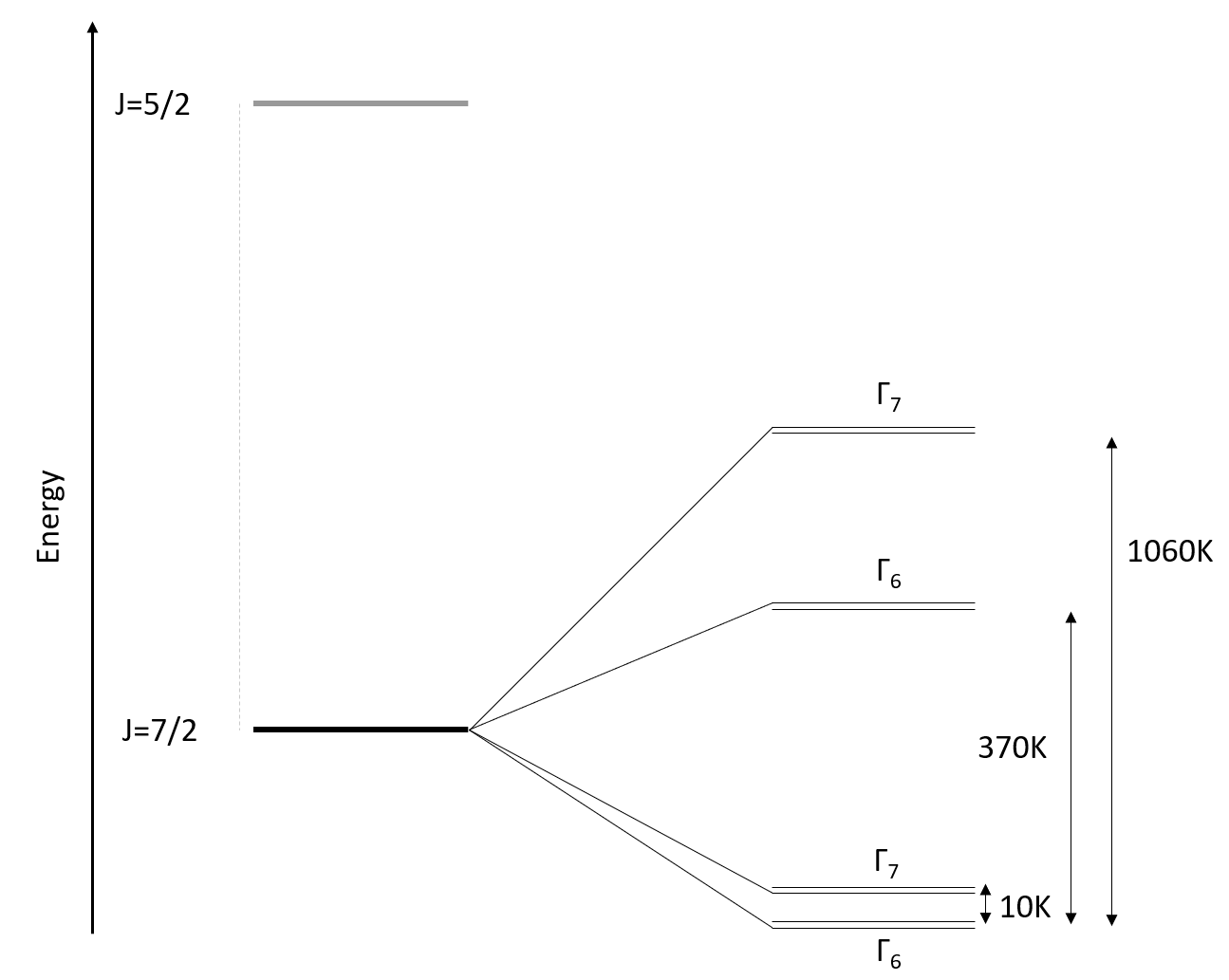}

\caption{\label{fig:S1} \textbf{YbRu$_2$Ge$_2$ CEF spectrum}  Spin-orbit coupling determines the ground state electronic mulitplet to have J=7/2, which is split by the surrounding crystalline potential to have 4 doublets, the lowest two in energy forming a quasi-quartet split by roughly 10K.}
\end{figure}

Where $O_l^m$ are the conventional Steven's operators \cite{Morinreview} and $B_l^m$ are coefficients to be determined. 
The resulting energy spectrum comprises 4 Kramers doublets (two doublets with $\Gamma_6$ symmetry, and two doublets with $\Gamma_7$ symmetry) and has been characterized by a combination of inelastic neutron scattering \cite{MuonYRG} and thermodynamic probes \cite{YRGCEF}.  The tentative spectrum of states proposed by Jeevan in \cite{Thesis} is $\Gamma_6$ at 91meV, $\Gamma_7$ at 32meV, $\Gamma_7$ at 0.9meV, and $\Gamma_6$ at 0meV (ground state), illustrated in Fig.~\ref{fig:S1}.
This is the spectrum that we use to calculate the low-temperature quadrupole strain susceptibility, as described in the main text and below in Appendix~\ref{section:S3}.

\section{X-ray diffraction data for YbRu$_2$Ge$_2$}
\label{section:S2}
\begin{figure}[H]
\centering
\includegraphics[width=0.5\linewidth]{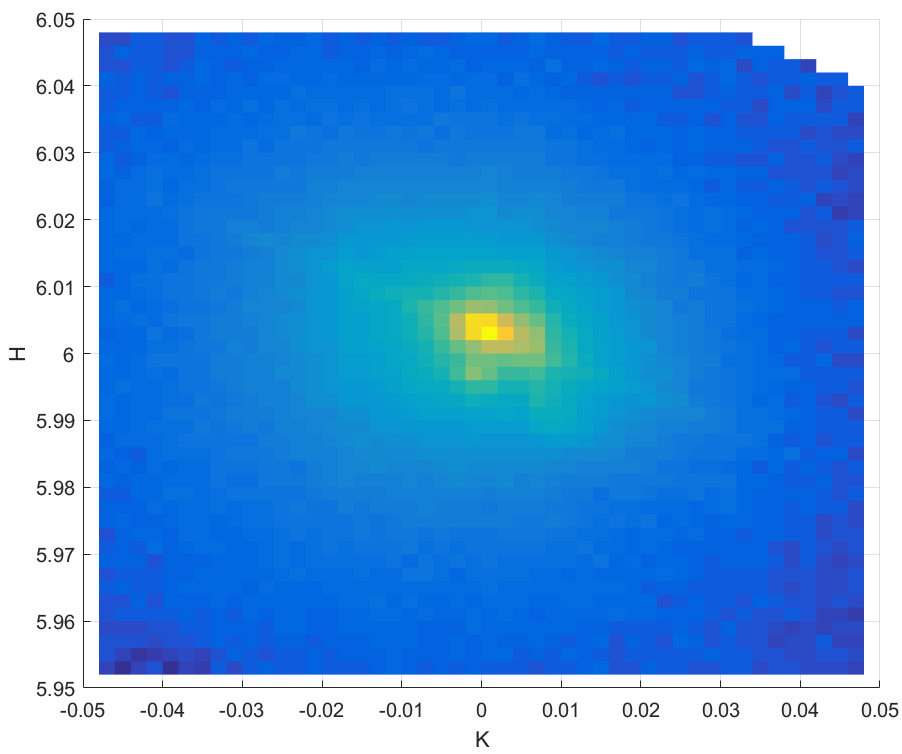}

\caption{\label{fig:S2} \textbf{Surface plot of log(intensity) at 12.2K} The material is still clearly tetragonal here, although it is displaying some spread in the momentum space direction that orthrorhombic domains are expected, possibly indicating critical fluctuations or a static response to unintentional strains from securing the crystal to the sample holder.}
\end{figure}

\begin{figure}[H]
\centering
\includegraphics[width=0.5\linewidth]{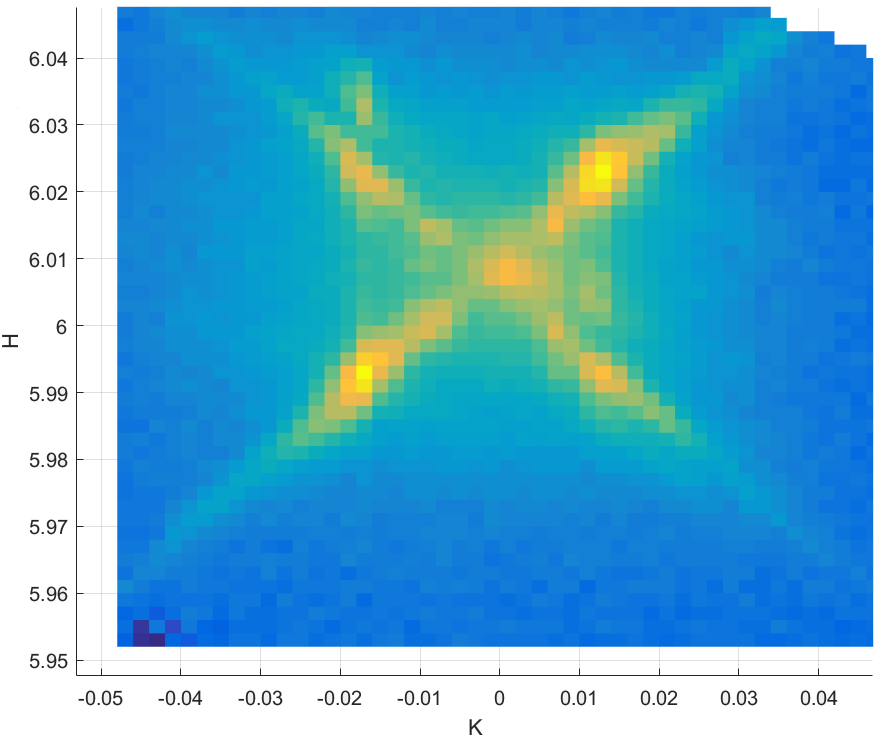}

\caption{\label{fig:S3}  \textbf{Surface plot of log(intensity) at 6.6K} The material is clearly orthrorhombic at this temperature, displaying multiple peaks.}
\end{figure}

\begin{figure}[H]
\centering
\includegraphics[width=0.8\linewidth]{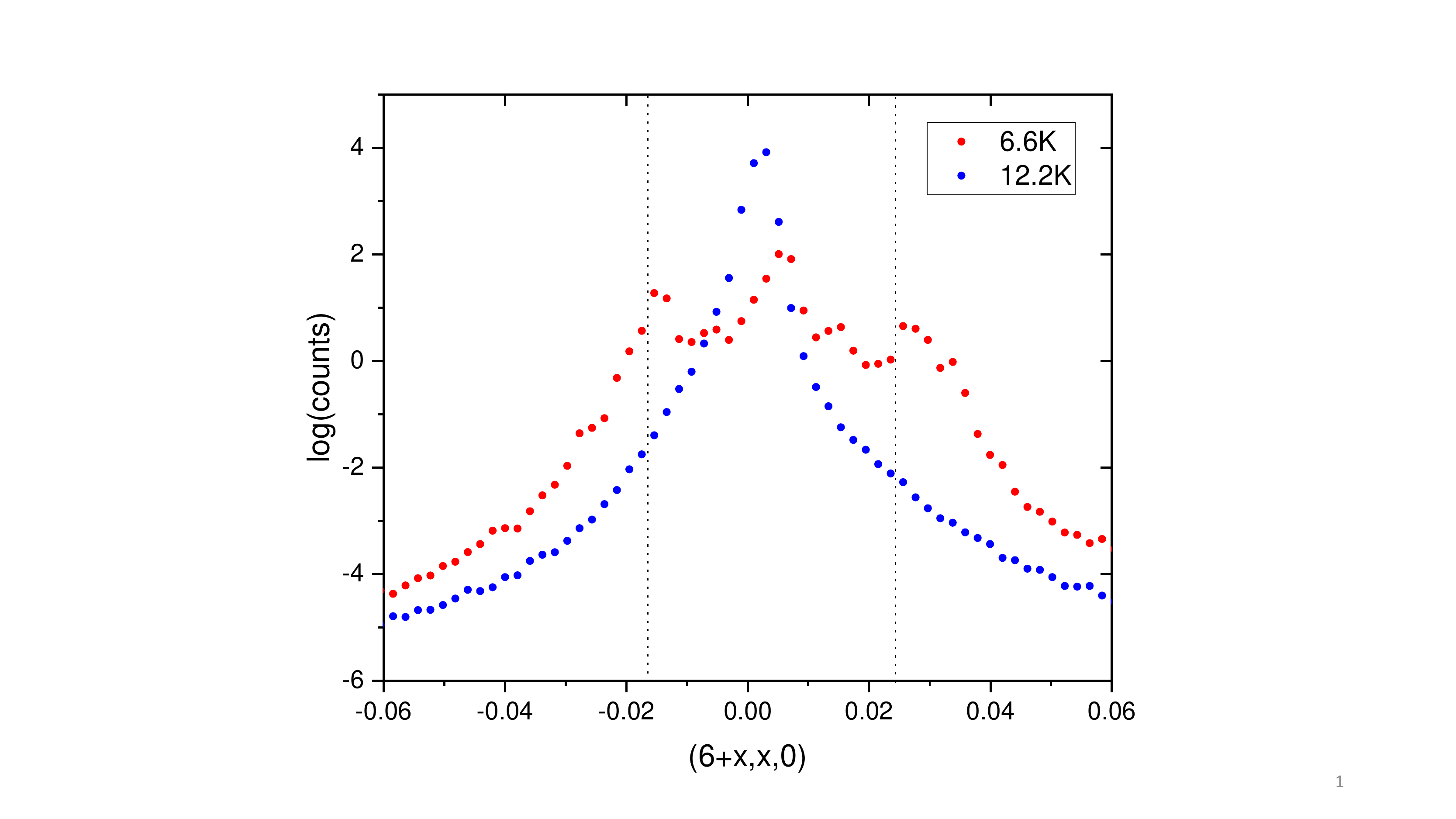}

\caption{\label{fig:S4} \textbf{Line cut of log(counts) along the (1,1,0) direction, centered at (6,0,0) } The dotted lines represent where the first moment of counts was determined to be, which was normalized by the lattice parameters to determine the orthrorhombic order parameter.}
\end{figure}

Low temperature, high-resolution, X-ray diffraction measurements were performed on beamline A2 at CHESS (Cornell High Energy Synchrotron Source). Splitting of the (6 0 0) Bragg peak was observed below 10.2K, with the new peaks indicative of an orthorhombic structural distortion with a B$_{1g}$ (x$^2$-y$^2$) symmetry, the associated domain structure of which results in 4 separate peaks along the (1 1 0) and (1 -1 0) directions \cite{Xraysplit}. Representative data are shown in Figures~\ref{fig:S2} and~\ref{fig:S3}, taken at 12.2 K (above T$_Q$) and 6.6 K (below T$_Q$) respectively. A line cut along the (1 1 0) direction for both data sets is shown in Figure~\ref{fig:S4}. The data in Figure~\ref{fig:S3} and~\ref{fig:S4} for $T < T_Q$ reveal the persistence of the central tetragonal peak, albeit with a reduced intensity, implying that some part of the illuminated volume of the crystal remains in the tetragonal state upon cooling through T$_Q$. Since the phase transition is characterized via heat capacity measurements to be continuous \cite{YRGCEF}, this observation implies heterogeneity of either the sample temperature or of the critical temperature T$_Q$. Thermodynamic and transport measurements indicate a maximum spread of critical temperatures of approximately 0.5 K  but at least in principle local strains due to sample mounting for the measurement can plausibly affect the critical temperature leading to a larger variation. Additional measurements would be necessary to characterize how rapidly T$_Q$ is affected by homogeneous strains of various symmetries in order to assess whether this is the origin of the effect. To best account for this when determining the orthorhombic order parameter (shown in Figure 1 of the main paper), we obtained the position of the first moment of counts along the (1 1 0) direction (above a background threshold), that were clearly not part of the original tetragonal peak.

\section{Quadrupole operators}
\label{section:S3}

The three relevant quadrupole operators, which correspond to the axis of the quadrupole oriented along [0 0 1] (z$^2$ symmetry), along [1 0 0] or [0 1 0] ($x^2-y^2$ symmetry), and along [1 1 0] or [1 -1 0] (xy symmetry) respectively, are given by the familiar Steven's operators:

\begin{equation}
O^0_2=3J_z^2-J(J+1)
\end{equation}
\begin{equation}
O^2_2=J_x^2-J_y^2=\frac{1}{2}(J_+^2 + J_-^2)
\end{equation}
\begin{equation}
P_{xy}=\frac{1}{2}(J_xJ_y+J_yJ_x)=\frac{-i}{4}(J_+^2 - J_-^2)
\end{equation}

In the restricted Hilbert space corresponding to the quasi-quartet CEF groundstate of YbRu$_2$Ge$_2$, these operators have the following matrix elements, where for convenience the basis of states is represented in the order (3/2, -1/2, -3/2, 1/2)

$$O^0_2=\begin{bmatrix} -15 & 0 & 0 & 0 \\  0 & -9 & 0 & 0\\ 0 & 0 & -15 & 0 \\ 0 & 0 & 0 & -9 \\ 
\end{bmatrix}$$

$$O^2_2=\begin{bmatrix} 0 & 2\sqrt{15} & 0 & 0 \\ 2\sqrt{15} & 0 &0 & 0\\ 0 &0& 0 & 2\sqrt{15}  \\ 0 & 0 & 2\sqrt{15}  & 0 \\ 
\end{bmatrix}$$

$$P_{xy}=\begin{bmatrix}  0 & -i\sqrt{15} & 0 & 0 \\ i\sqrt{15} & 0 &0 & 0\\ 0 &0& 0 & i\sqrt{15}  \\ 0 & 0 & -i\sqrt{15}  & 0 \\ 
\end{bmatrix}$$

Noticing the correspondence to the Pauli spin matrices, we introduce for completeness a third, octupole, operator, that will also have finite matrix elements in this basis: $O^{-2}_3 = \frac{-i}{4}(J_z(J_+^2 - J_-^2) + (J_+^2 - J_-^2) J_z)$

$$O^{-2}_3=\begin{bmatrix} 0 &- i\sqrt{15} & 0 & 0 \\ i\sqrt{15} & 0 &0 & 0\\ 0 &0& 0 &- i\sqrt{15}  \\ 0 & 0 & i\sqrt{15}  & 0 \\ 
\end{bmatrix}$$

Inspection of these expressions reveals they can be written as tensor products of the canonical Pauli matrices and the identity, I:

\begin{equation}
 O^2_0=I \otimes (-3\sigma_z -12I) 
\end{equation}

\begin{equation}
 O^2_2=I \otimes (2\sqrt{15}\sigma_x) 
\end{equation}

\begin{equation}
P_{xy}=\sigma_z \otimes (\sqrt{15}\sigma_y) 
\end{equation}

\begin{equation}
O^{-2}_3=I \otimes (\sqrt{15}\sigma_y) 
\end{equation}

With reference to the tensor product formula:
\begin{equation}
[I\otimes A, I\otimes B]=I\otimes [A,B]
\end{equation}
and noting that constants don't affect commutation relations,  the three operators $ O^0_2$, $O^2_2$, and  $O^{-2}_{3}$ obey the canonical commutation relations.  The quasi-quartet ground state can be thought of as two replicas of a pseudo-spin $\frac{1}{2}$ doublet, where the two replicas arise as a consequence of Kramer's theorem. These three operators will then serve as the effective spin operators in the three spatial dimensions of the pseudo-spin space.  The quartet is split (by roughly $\Delta_0=10K$) due to the tetragonal point symmetry of the CEF, yielding a finite $\sigma_z$ (i.e a finite $O^0_2$ quadrupole moment) above T$_Q$. Mixing of these eigenstates, as described in Table 1 of the main text, can then yield finite quadrupole moments O$_2^2$ or P$_{xy}$.


\section{Quadrupole-strain Susceptibility}
\label{section:S4}
Externally applied stresses cause finite strains, which in turn affect the eigenstates and eigenvalues of H$_{CEF}$, shifting and admixing the states described in Appendix~\ref{section:S1}. The magneto-elastic coupling (MEC) Hamiltonian is given by
$$H=H_{CEF}+  \sum_{\Gamma_i}{B^l_m\varepsilon_{\Gamma_i}Q_{\Gamma_i}}$$

Where B$^l_m$ are coefficients yet to be determined, and $\Gamma_i$ are irreducible representations of the point group.
Applying a non-zero stress which induces a strain $ \left(\varepsilon_{\gamma}=\varepsilon_{xx}-\varepsilon_{yy}\right)$ will induce a finite moment of $\langle O^2_2 \rangle$, which will perturbatively change the existing Hamiltonian (in the basis of the quasi-quartet) to be in the form:

$$H=H_{CEF}+  B^2_2\varepsilon_{\gamma}O^2_2= I \otimes\begin{bmatrix} \Delta_0/2 &2\sqrt{15}B^2_2\varepsilon_{\gamma} \\2\sqrt{15}B^2_2\varepsilon_{\gamma} & -\Delta_0/2 \\
\end{bmatrix} $$

Diagonalizing this matrix gives a new energy gap of:
\begin{equation}
\Delta/2=\sqrt{(\Delta_0/2)^2 + 60(B_2^2\varepsilon_{\gamma})^2} 
\end{equation}
 
The thermal expectation value of the quadrupolar moment $\langle O^2_2 \rangle$ is now:

\begin{equation}
 \langle O^2_2 \rangle=\frac{120B_2^2\varepsilon_{\gamma}}{\Delta}\tanh{\Bigg(\frac{\Delta}{2T}\Bigg)}
\end{equation}

Thus the quadrupole-strain susceptibility is:

\begin{equation}
\chi_{Q_{B_{1g}}}=\frac{dQ_{B_{1g}}}{d\varepsilon_{B_{1g}}}\bigg|_{\varepsilon\to 0}= \frac{60B_2^2}{T}
\end{equation}
 when $T>>\Delta_0$ , and

\begin{equation}
\chi_{Q_{B_{1g}}}=\frac{dQ_{B_{1g}}}{d\varepsilon_{B_{1g}}}\bigg|_{\varepsilon\to 0}= \frac{120B_2^2}{\Delta_0}\tanh{\bigg(\frac{\Delta_0}{2T}\bigg)}
\end{equation}
in general.

Although in the case of the B$_{2g}$ order parameter $P_{xy}$ the Hamiltonian cannot be written as concisely, a similar result is still obtained:

\begin{equation}
\chi_{Q_{B_{2g}}}=\frac{dQ_{B_{2g}}}{d\varepsilon_{B_{2g}}}\bigg|_{\varepsilon\to 0}= \frac{15B_{xy}}{T}
\end{equation}
 when $T>>\Delta_0$ , and

\begin{equation}
\chi_{Q_{B_{2g}}}=\frac{dQ_{B_{2g}}}{d\varepsilon_{B_{2g}}}\bigg|_{\varepsilon\to 0}= \frac{30B_{xy}}{\Delta_0}\tanh{\bigg(\frac{\Delta_0}{2T}\bigg)}
\end{equation}
in general.

Hence both the B$_{1g}$ and B$_{2g}$ quadrupole strain susceptilities can be put in the form:
\begin{equation}
\chi_{Q_{\Gamma_i}}=\frac{dQ_{\Gamma_i}}{d\varepsilon_{\Gamma_i}}\bigg|_{\varepsilon\to 0}= \frac{2\langle Q_{\Gamma_i}\rangle_{o}^2 B_{\Gamma_i}}{\Delta_0}\tanh{\bigg(\frac{\Delta_0}{2T}\bigg)}
\end{equation}

Assuming no B$_{1g}$ or B$_{2g}$ strain is being applied, the A$_{1g}$ quadrupole-strain susceptibility is:

\begin{equation}
\chi_{Q_{A_{1g}}}=\frac{dQ_{A_{1g}}}{d\varepsilon_{A_{1g}}}\bigg|_{\varepsilon\to 0}=\frac{\langle Q_{A_{1g}}\rangle_{o}^2(B^0_2)_1 sech^2{\big(\frac{\Delta_0}{2T}}\big)}{T}
\end{equation}
 Where $\langle Q_{A_{1g}} \rangle_{o} =3$

Adding in mean-field interactions of the form $H_{QQ}=K_{\Gamma_i}\langle Q_{\Gamma_i} \rangle Q_{\Gamma_i}$ will renormalize the quadrupole-strain susceptibilities to be in the form:

\begin{equation}
\chi_{Q_{\Gamma_i}}=\frac{dQ_{\Gamma_i}}{d\varepsilon_{\Gamma_i}}\bigg|_{\varepsilon\to 0}= \frac{2 B_{\Gamma_i}\langle Q_{\Gamma_i}\rangle_{o}^2\tanh{\bigg(\frac{\Delta_0}{2T}\bigg)}}{\Delta_0-2K_{\Gamma_i}\langle Q_{\Gamma_i}\rangle_{o}^2\tanh{\bigg(\frac{\Delta_0}{2T}\bigg)}}
\end{equation}

For $ T>>\Delta/2$ this becomes the familiar expression:

\begin{equation}
\chi_{Q_{\Gamma_i}}=\frac{B_{\Gamma_i}\langle Q_{\Gamma_i}\rangle_{o}^2}{T-\langle Q_{\Gamma_i}\rangle_{o}^2 K_{\Gamma_i}}
\end{equation}

For $A_{1g}$ this renormalizes it to become:
\begin{equation}
\chi_{A1g}=\frac{B_0^2\langle Q_2^0\rangle_{o}^2 sech^2{\big(\beta \big(\Delta/2+3K^0_2\langle O^0_2 \rangle|_{\varepsilon=0}\big)\big)}}{T-\langle Q_2^0\rangle_{o}^2K^0_2 sech^2{\big(\beta \big(\Delta/2+3K^0_2\langle O^0_2 \rangle_{\varepsilon=0}\big)\big)}}
\end{equation}

The proposed CEF quasi-quartet states \cite{Thesis} can be substituted in to find the actual values : $\langle Q_i \rangle_o$:
$\langle O^0_2 \rangle_o =3.035$,
$\langle O^2_2 \rangle_o =8.3185$, and
$\langle P_{xy} \rangle_o =3.4660$

With these values the elastoresistivity measurements can be fit to to obtain absolute values for $K_i$ and the gap $\Delta$ and relative ratios of $B_i$.


\section{Relation of Elastoresistivity to Quadrupole-strain Susceptibility}
\label{section:S5}
To show how the proportionality between the quadrupole-strain susceptibilities and the elastoresistivity coefficients is obtained, we follow Friederich and Fert \cite{Aniso} and extend their argument to tetragonal systems, replacing the magnetic field with strain as the source of the quadrupole moment.  If we make the following assumptions that: a) the strain is perturbative, hence the quadrupole moments can be treated as impurities but the system has a infinitesimal overall quadrupole moment;  b) The scattering is dominated by isotropic (in the ab plane) elastic scattering potentials V$\delta(r_i)$ at each Yb site i; c) We can use the first Born approximation to obtain the scattering rate $W_{kk'}$; and d) The conduction electrons are primarily s-wave and p-wave in character, then we can follow the argument laid out in ref \cite{Aniso}.

We begin by writing down the scattering interaction between s and p wave conduction electrons and $4f$ sites originally derived by Kondo:

\begin{equation}
V_{scatt}=\sum_{kk'}{\bigg[V-\frac{D}{k^2_f}\bigg ((J\cdot k)(J\cdot k')-\frac{J(J+1)}{3}k\cdot k'\bigg)\bigg]a_{k'}^\dagger a_k'}
\end{equation}
Where D is the coefficient of the quadrupolar scattering potential from the $4f$ electrons, and V in the sum is the previously mentioned strength of the isotropic scattering potential.
When this potential is plugged into Fermi's Golden rule, assuming the quadrupole term is perturbatively small, the anisotropic cross terms lead to a resistivity ratio directly from Ref. \cite{Aniso}, Equation 3:

\begin{equation}
\frac{\rho^Q_i}{\rho_0}=\frac{2D}{3V}\bigg(\langle J_i^2 \rangle -\frac{J(J+1)}{3}\bigg)
\end{equation}

Where i is the direction of the current, and $\rho_0$ is the resistivity due to the isotropic scattering potential (isotropic only in the ab plane in the case of YbRu$_2$Ge$_2$)

Thus:
\begin{equation}
\frac{\rho^Q_x -\rho^Q_y}{2\rho_0}=\frac{D}{3V}\bigg(\langle J_x^2 \rangle -\langle J_y^2 \rangle\bigg)=\frac{D}{3V}\langle O^2_2 \rangle
\end{equation}

Because inelastic scattering should only be dependent on the magnitude of the gap and matrix elements like $\langle | Q_i | \rangle^2$, its $B_{1g}$ component induced by strain should be close to zero. Similarly the anisotropic part of the Kondo scattering should be close to zero, as there is no reason to suggest there are quadrupolar aspects of the coupling of conduction electrons to the magnetic aspects of the $4f$ sites. Thus, taking the appropriate strain derivatives (in this case, with respect to $\varepsilon_{xx}-\varepsilon_{yy}$), we find that the elastoresistivity associated with scattering from the $4f$ orbital is directly proportional (with a temperature independent proportionality coefficient) to the B$_{1g}$ quadrupole-strain susceptibility.

\begin{equation}
\label{equation:S24}
\cfrac{\cfrac{\partial (\rho^{xx}_{4f} -\rho^{yy}_{4f})}{\rho_{4f}^0}}{\partial\varepsilon_{B1g}}\Bigg|_{\varepsilon\to 0}\approx \cfrac{D}{3V}\frac{\partial\langle O^2_2 \rangle}{\partial\varepsilon_{B1g}} \propto \chi_{B1g}
\end{equation}

Several scattering processes contribute to the resistivity of YbRu$_2$Ge$_2$. Assuming validity of Matthiessen's rule,
$\rho_{YbRu2Ge2} = \rho_{imp} + \rho_{e-ph} + \rho_{e-e} + \rho_{4f}$
where $\rho_{imp}$ arises from impurity scattering, $\rho_{e-ph}$ from electron-phonon interactions, $\rho_{e-e}$ from electron-electron scattering, and $\rho_{4f}$ is defined above. At least in principle, each of these terms can have an associated elastoresistivity; the expression derived in Eq.~\ref{equation:S24} above relates only to the $4f$ part. Contributions to the resistivity and elastoresistivity arising from $\rho_{imp}$, $\rho_{e-ph}$ and $\rho_{e-e}$ can be subtracted by considering a non-magnetic analog that has the same crystal structure, the same band structure and a similar impurity concentration. YRu$_2$Ge$_2$ (note that Y = Yttrium, different to Ytterbium Yb) potentially provides such a non-magnetic analog. Such a subtraction would then yield,

\begin{equation}
\cfrac{\cfrac{\partial (\rho^{xx}_{4f} -\rho^{yy}_{4f})}{\rho_{4f}^0}}{\partial\varepsilon_{B1g}}\approx \cfrac{\cfrac{\partial (\rho^{xx}_{YbRu_2Ge_2} -\rho^{yy}_{YbRu_2Ge_2})}{\rho_{YbRu_2Ge_2}^0-\rho^0_{YRu_2Ge_2}} -\cfrac{\partial (\rho^{xx}_{YRu_2Ge_2} -\rho^{yy}_{YRu_2Ge_2})}{\rho_{YbRu_2Ge_2}^0-\rho^0_{YRu_2Ge_2}}}{\partial\varepsilon_{B1g}}
\end{equation}

where superscripts `0' refer to zero strain conditions.

The unstrained resistivity of YRu$_2$Ge$_2$, $\rho_{YRu2Ge2}^0$, is found to be almost an order of magnitude smaller than that of YbRu$_2$Ge$_2$ (see Figure 3 in the main paper). Furthermore, normal metals far from any electronic instabilities, typically exhibit very small elastoresistivities. Hence, we can safely make the approximation that:

\begin{equation}
\cfrac{\cfrac{\partial (\rho^{xx}_{4f} -\rho^{yy}_{4f})}{\rho_{4f}^0}}{\partial\varepsilon_{B1g}}\approx \cfrac{\cfrac{\partial (\rho^{xx}_{YbRu_2Ge_2} -\rho^{yy}_{YbRu_2Ge_2})}{\rho_{YbRu_2Ge_2}^0}}{\partial\varepsilon_{B1g}}\propto \chi_{B1g}
\end{equation}

Hence elastoresistivity will be a direct measure of the quadrupole-strain susceptibility given these conditions.

\section{Linearity of the B$_{1g}$ elastoresistivity}
\label{section:S6}
\begin{figure}[H]
\centering
\includegraphics[width=0.5\linewidth]{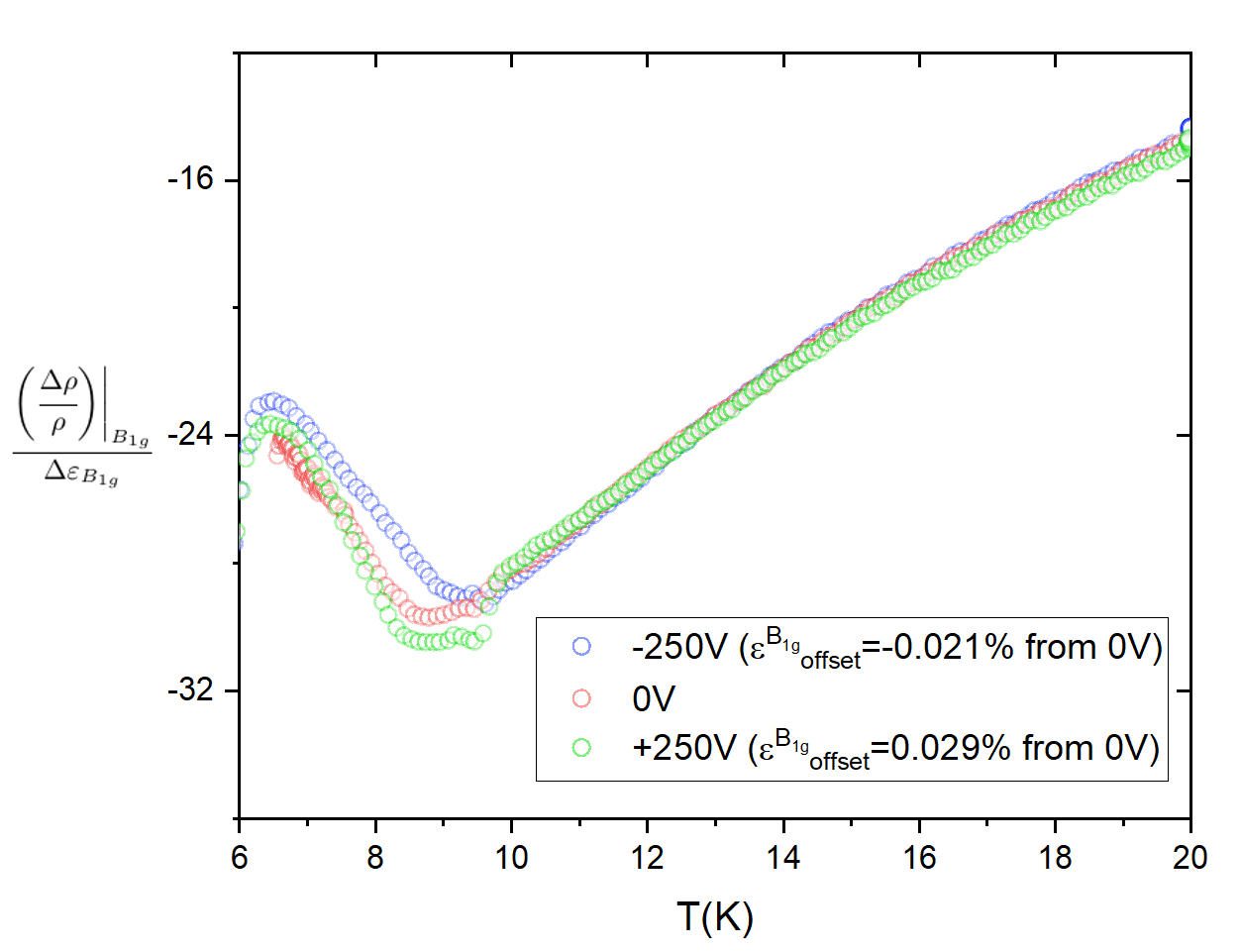}

\caption{\label{fig:S5} \textbf{Strain dependence of the B$_{1g}$ elastoresistivity response} The B$_{1g}$ quadrupole-strain susceptibility displays little sensitivity to tuning B$_{1g}$ offset strains above the quadrupolar phase transition indicating this channel is dominated by linear behavior, and hence $\big(\frac{\Delta \rho}{\rho}\big)_{B_{1g}}/\Delta \varepsilon_{B_{1g}}$ provides a good measure of the linear elastoresistivity coefficients for this symmetry channel, $m_{11}$-$m_{12}$}
\end{figure}

An effective way to show that the elastoresistivity is linear in strain while using the AC technique that we describe in the main paper is to perform these measurements for a variety of offset bias strains. In Fig.~\ref{fig:S5} we show data for the B$_{1g}$ response for measurements performed with an AC amplitude corresponding to a peak-to-peak voltage applied to the PZT stack of 40 volts, with a simultaneous DC bias voltage of 0V, -250V and + 250V in the temperature range from 6 to 20 K. Over this temperature range, these offset voltages correspond to DC strain offsets (relative to 0V) of approximately 0, -0.021\%, and 0.029\% .  As can be seen by inspecting the figures, in the temperature range above T$_Q$, the data almost perfectly line up, demonstrating the absence of any significant non-linear response.

\section{Non-linearity of the A$_{1g}$ elastoresistivity}
\label{section:S7}
\begin{figure}[H]
\centering
\includegraphics[width=0.5\linewidth]{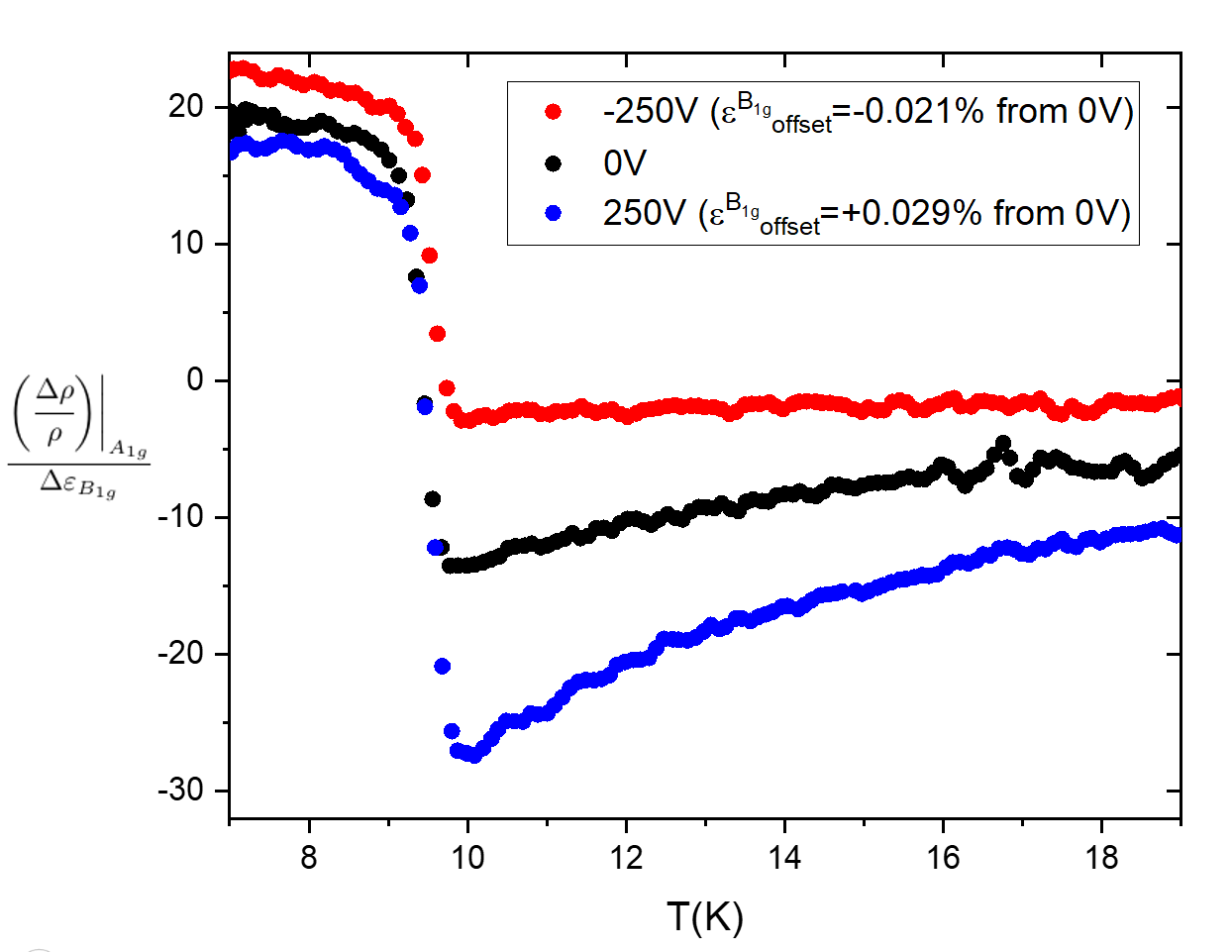}

\caption{\label{fig:S6} \textbf{Non-linearity of the A$_{1g}$ response with respect to B$_{1g}$ strain.} The A$_{1g}$ quadrupole-strain susceptibility displays striking sensitivity to tuning offset B$_{1g}$ strains above the quadrupolar phase transition.}
\end{figure}

\begin{figure}[H]
\centering
\includegraphics[width=0.5\linewidth]{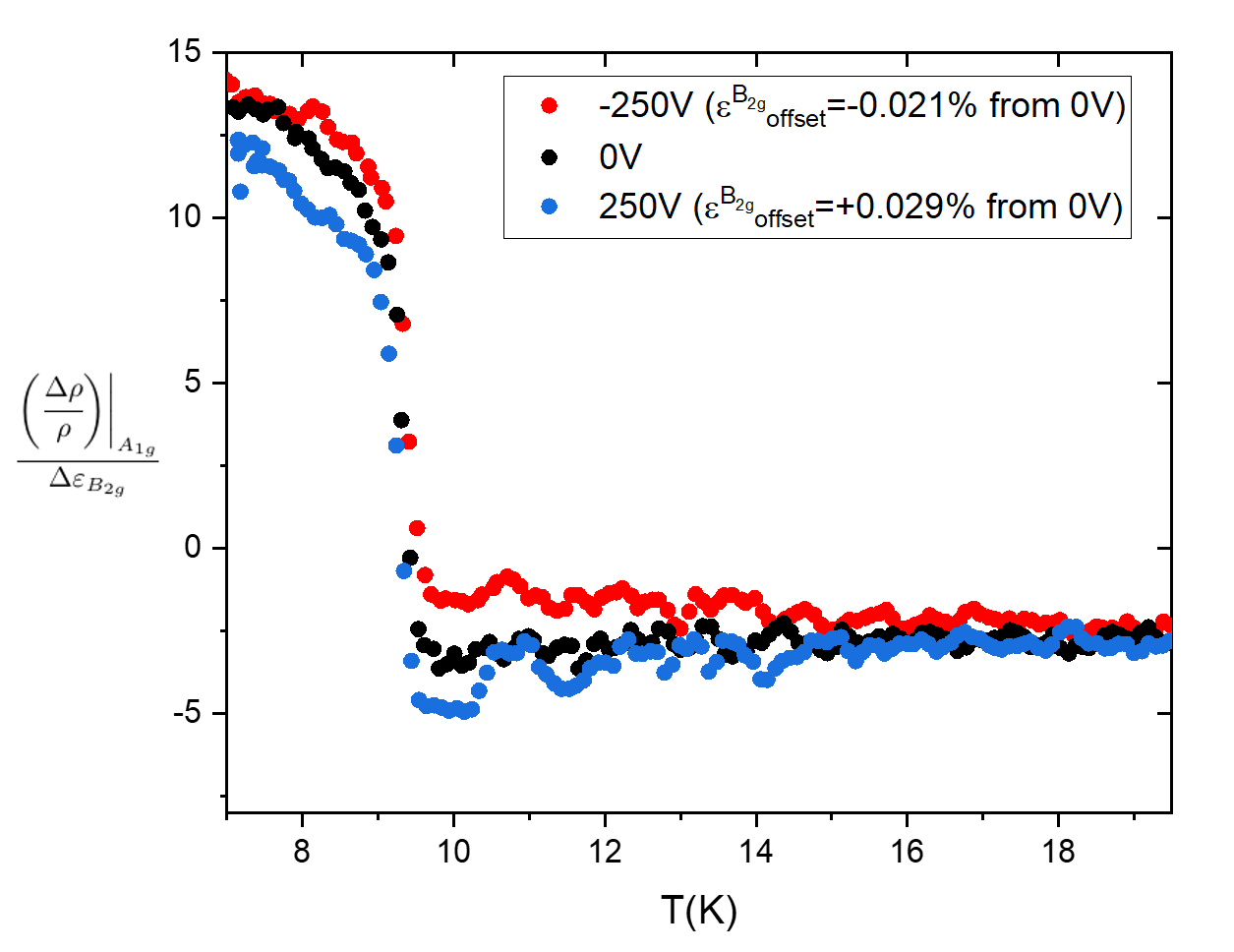}

\caption{\label{fig:S7} \textbf{Non-linearity of the A$_{1g}$ response with respect to B$_{2g}$ strain.} The A$_{1g}$ quadrupole-strain susceptibility displays small sensitivity to tuning offset A$_{1g}$ and B$_{2g}$ strains above the quadrupolar phase transition}
\end{figure}

In contrast to the case of B$_{1g}$ and B$_{2g}$ response, the in-plane A$_{1g}$ response $\frac{\rho_{xx} +  \rho_{yy}}{\rho_0}$ exhibited a striking non-linearity. Fig.~\ref{fig:S6} shows the response for measurements performed with an AC amplitude corresponding to a peak-to-peak voltage applied to the PZT stack of 40 volts, with a simultaneous DC bias voltage of 0V, -250V and + 250V in the temperature range from 6 to 20 K. Over this temperature range, these offset voltages correspond to DC strain offsets (relative to 0V) of approximately 0, -0.021\%, and 0.029\% . The sample was oriented on the PZT stack such that the crystal experienced a combination of A$_{1g}$ and B$_{1g}$ strains for Fig.~\ref{fig:S6}, and A$_{1g}$ and B$_{2g}$ strains for Fig.~\ref{fig:S7}. As can be seen, there is a striking difference between the measurements, indicating the presence of a substantial non-linear A$_{1g}$ elastoresistivity in response to B$_{1g}$ symmetry strains.

Although the A$_{1g}$ response is not the main subject of this paper, we outline below the origin of this effect, making particular reference to how this shows up in an AC measurement.

To second order in strain, a crystal that experiences both A$_{1g}$ and B$_{1g}$ symmetry strains can experience an A$_{1g}$ elastoresistance response given by

\begin{equation}
\label{equation:S26}
\big( \frac{\Delta \rho}{\rho_0}\big)_{A_{1g}}=m^{A_{1g}}_{A_{1g}}\varepsilon_{A_{1g}} + m^{A_{1g}, A_{1g}}_{A_{1g}}[\varepsilon_{A_{1g}}]^2 
+ m^{B_{1g}, B_{1g}}_{A_{1g}}[\varepsilon_{B_{1g}}]^2 
\end{equation}

Where we are following the notation used in Ref. [\cite{Nonlin}]. The linear term proportional to $\varepsilon_{A_{1g}}$ is allowed by symmetry as well as quadratic terms proportional to $\varepsilon_{B_{1g}}^2$.

Since the AC Elastoresistivity method used for these measurements locks in to the response at the frequency $\omega$ at which the strain is applied, it is useful to write the strains out as a combination of DC offset strains (arising for example from thermal expansion mismatches and glue strains, as well as intentional bias strains as mentioned previously) and AC applied strains:

\begin{equation}
\varepsilon^{tot}_{i}=\varepsilon^{DC}_{i}+\varepsilon^{AC}_{i}\cos{\omega t}
\end{equation}

Where i represents the symmetry channel and the amplitude of the AC term depends on the voltage waveform applied to the piezo.  Substituting into Eq.~\ref{equation:S26} and focusing on the amplitude of the signal that will be locked into at frequency $\omega$:

\begin{equation}
\big( \frac{\Delta \rho}{\rho_0}\big)^{AC}_{A_{1g}}=m^{A_{1g}}_{A_{1g}}\varepsilon^{AC}_{A_{1g}} + 2m^{A_{1g}, A_{1g}}_{A_{1g}}\varepsilon^{DC}_{A_{1g}}\varepsilon^{AC}_{A_{1g}}
+ 2m^{B_{1g}, B_{1g}}_{A_{1g}}\varepsilon^{DC}_{B_{1g}}\varepsilon^{AC}_{B_{1g}} 
\end{equation}

Thus we can expect both a linear and non-linear contribution to the signal, with the strength of the non-linear part determined by both the amount of offset strain and the quadratic coefficient of that channel. 

A similar difference in non-linear elastoresistivity coefficients was recently observed  for the underdoped Fe-based superconductor, Ba(Fe$_{0.975}$Co$_{0.025}$)$_2$As$_2$ \cite{Nonlin}. In that material, the nematic transition occurs in the B$_{2g}$ symmetry channel and the quantity m$^{B_{2g},B_{2g}}_{A_{1g}}$ exhibits a divergence. In the present case, YbRu$_2$Ge$_2$ undergoes a nematic transition in the B$_{1g}$ symmetry channel, and the quantity m$^{B_{1g},B_{1g}}_{A_{1g}}$ appears to grow very large. Both results highlight the role played by nematic (quadrupole) fluctuations in affecting the isotropic properties of materials . In the case of YbRu$_2$Ge$_2$, the observation of a large m$^{B_{1g},B_{1g}}_{A_{1g}}$ adds further evidence to our conclusion that the quadrupole-strain susceptibility is large in the B$_{1g}$ channel but small in the B$_{2g}$ channel.

\section{Focused Ion Beam parameters}
\label{section:S8}
The instrument used to etch the samples was an FEI Helios NanoLab 600i DualBeam FIB/SEM, containing both a focused Ga+ ion beam ("Tomahawk") and a high resolution field emission scanning electron ("Elstar") column. Combined with advances in patterning, scripting, and a suite of accessories, these features make milling, imaging, analysis, and sample preparation down to the nanoscale possible.

A 65 nA ion current was used to etch through the samples. Gallium implantation is expected to affect a depth of less than 100 nm from the surface roughly, which is inconsequential for the bulk resistivity measurements that were performed.
\section{Crystal Growth}
\label{section:S9}
Single crystals of YbRu$_2$Ge$_2$ were grown using an unseeded flux method, with Indium being the flux and the other precursors added in stoichiometrically.
The flux ratio was varied from 96-98\%, with 97.5\% found to produce both the largest size individual crystals and also the greatest yield. To help ensure inclusion of the high-melting point Ru into the melt, the elemental Ru and Ge precursors were arc-melted in a mono-arc furnace. The elements were then combined into an alumina crucible, which was sealed inside a Ta crucible to prevent oxidation and to contain the flux. The crucibles were heated to a max temperature of 1450K for 6-12 hours, and then cooled to 1200K at approximately 4K/hour. The alumina crucibles were then sealed quartz and spun in a centrifuge at 400K to remove the Indium flux from the crystals  The resulting crystals were etched in HCl acid for several months until they were easily cleaveable.

%
%
%
%

\end{document}